# Coexistence of bulk superconductivity and ferromagnetism in $CeO_{1-x}F_xBiS_2$

Satoshi Demura,[1,2] Keita Deguchi,[1,2] Yoshikazu Mizuguchi,[3] Kazuki Sato,[1] Ryouta Honjyo,[1] Aichi Yamashita,[1] Takuma Yamaki,[1,2] Hiroshi Hara,[1,2] Tohru Watanabe,[1,2] Saleem J. Denholme,[1] Masaya Fujioka,[1] Hiroyuki Okazaki,[1] Toshinori Ozaki,[1] Osuke Miura,[3] Takahide Yamaguchi,[1] Hiroyuki Takeya,[1] and Yoshihiko Takano[1,2]

E-mail : Demura.Satoshi@nims.go.jp

[1]National Institute for Materials Science, 1-2-1 Sengen, Tsukuba, Ibaraki 305-0047, Japan

[2]University of Tsukuba, Graduate School of Pure and Applied Sciences, Tsukuba, Ibaraki 305-8577, Japan

[3]Tokyo Metropolitan University, Hachioji, Tokyo 192-0397, Japan

**Abstract**

We show the observation of the coexistence of bulk superconductivity and ferromagnetism in $CeO_{1-x}F_xBiS_2$($x$ = 0 - 1.0) prepared by annealing under high-pressure. In $CeO_{1-x}F_xBiS_2$ system, both superconductivity and two types of ferromagnetism with respective magnetic transition temperatures of 4.5 K and 7.5 K are induced upon systematic F substitution. This fact suggests that carriers generated by the substitution of O by F are supplied to not only the $BiS_2$ superconducting layers but also the CeO blocking layers. Furthermore, the highest superconducting transition temperature is observed when the ferromagnetism is also enhanced, which implies that superconductivity and ferromagnetism are linked to each other in the $CeO_{1-x}F_xBiS_2$ system.

## Introduction

Layered superconductors are attractive materials due to the discovery of superconductors with high transition temperatures and unconventional mechanisms behind their superconductivity, which cannot be explained by the conventional BCS theory [1-13]. Furthermore, some compounds show superconductivity which coexist with long range magnetic ordering [14-28]. Since long range magnetic ordering usually competes against superconductivity, the materials possessing the coexistence of superconductivity and magnetism are of great interest for basic physics and applications.

Recently, new layered superconductors with superconducting $BiS_2$ layers intermixed with blocking layers were discovered [29-40]. As-grown $LaO_{0.5}F_{0.5}BiS_2$ is a typical $BiS_2$-based superconductor and shows superconductivity with a temperature ($T_c$) of 3 K. The $T_c$ increases up to 10 K by annealing under high-pressure. This fact indicates that the optimization of local crystal structure is important for superconductivity with a higher transition temperature in the $BiS_2$-based superconductors [41-47]. Another typical $BiS_2$-based material is $CeO_{1-x}F_xBiS_2$, which shows weak signal of superconductivity around 3 K while exhibiting ferromagnetism in the CeO blocking layers at low temperature. This result suggests that ferromagnetism and superconductivity could coexist in this system. If bulk superconductivity of the $BiS_2$ layers could be induced while keeping the ferromagnetic ordering in the CeO layers, $CeO_{1-x}F_xBiS_2$ could become one of the potential materials which achieves the coexistence of ferromagnetism and superconductivity.

Here, we show the observation of the coexistence of ferromagnetism and bulk superconductivity in $CeO_{1-x}F_xBiS_2$ annealed under high-pressure. Furthermore, the established phase diagram suggests that superconductivity in the $BiS_2$ layers and ferromagnetism in the CeO layers are linked to each other.

## Experimental

The polycrystalline samples of $CeO_{1-x}F_xBiS_2$ ($x$ = 0 - 1.0) were prepared with a solid-state reaction and then annealed under high-pressure using a Cubic-Anvil-type high-pressure machine. $Bi_2S_3$ powders were obtained by sintering the mixtures of Bi grains and S grains in the evacuated quartz tube at 500 °C for 10 hours. Mixtures of Bi grains, $Bi_2S_3$ grains, $Bi_2O_3$ powders, $BiF_3$ powders and $Ce_2S_3$ powders with nominal compositions of $CeO_{1-x}F_xBiS_2$($x$ = 0.0 - 1.0) were ground, pelletized, and sealed into an evacuated quartz tube. The tube was heated at 800 °C for 10 hours. The obtained pellets were ground and annealed at 600 °C for 1 hour under a hydrostatic pressure of 3 GPa. We denote as-grown samples as AG, and high pressure samples as HP in this paper. The obtained samples were characterized by X-ray diffraction with Cu-K$\alpha$ radiation using the $\theta$-$2\theta$ method. The temperature dependence of magnetic susceptibility was measured by a superconducting quantum interface device (SQUID) magnetometer with an applied field of 10 Oe. The resistivity measurements were performed using the four-terminal method from 300 to 2 K.

**Results**

Figure 1(a) and (b) show the powder X-ray diffraction profile for the as-grown (AG) and high-pressure (HP) samples of $CeO_{1-x}F_xBiS_2$ ($x$ = 0.0 - 1.0), respectively. Almost all of the peaks are indexed to the space group *P*4/*nmm* for both the AG and HP samples. Impurity phases of $Bi_2S_3$ appear above $x$ = 0.7. Enlarged XRD figures near the (110), (114) and (200) peaks for the AG samples are shown in Fig.1(c). These each present themselves as a single peak between $x$ = 0.0 and 0.6. However, other peaks appear above $x$ = 0.7 and become stronger with increasing F concentration. Furthermore, the peaks of the tetragonal phase seem to shift with increasing F concentration. These facts imply that another phase appears or the tetragonal structure is strained above $x$ = 0.7. Tomita *et al.* reported a structural transition from tetragonal to monoclinic in $LaO_{0.5}F_{0.5}BiS_2$ under high-pressure [46]. Having considered the fact in $LaO_{0.5}F_{0.5}BiS_2$, structural transition or anisotropic strain could occur with decreasing lattice volume in the $CeO_{1-x}F_xBiS_2$ system as well. The observed peaks of $CeO_{1-x}F_xBiS_2$ samples may partly begin to be strained into a lower-symmetry phase due to lattice contraction upon F substitution. The lattice constants $a$ and $c$ of the AG and HP samples are plotted against the nominal $x$ dependence in Fig. 1(d) and (e). These lattice constants were estimated from the $2\theta$ values and the corresponding Miller indices. For the as-grown samples, the $a$ lattice constant increases with increasing F concentration between $x$

= 0.0 and 0.6, while they slightly decrease above $x$ = 0.7. The $c$ lattice constant markedly decreases between $x$ = 0.0 and 0.6 and keeps an almost constant value from $x$ = 0.7 to 1.0. This decrease in the $c$ lattice constant between $x$ = 0.0 and 0.6 indicates that F substitutes O as expected. After high pressure annealing, the $a$ and $c$ lattice constant exhibit almost the same behavior as the AG samples for all F concentrations. Therefore, the change of the crystal structure before and after high-pressure annealing is not large, but some local structures should be tuned because superconducting properties significantly change after high-pressure annealing.

The temperature dependence of resistivity for AG samples of $CeO_{1-x}F_xBiS_2$ ($x$ = 0 - 1.0) are shown in Fig. 2(a) and (b). The resistivity for all F-doped samples increases with decreasing temperature. This semiconducting behavior in the normal-state region is the same as that of $LnO_{1-x}F_xBiS_2$ (Ln=La, Nd, Pr, Nd, Yb) and $Sr_{1-x}La_xFBiS_2$ [30-37]. The low temperature region is enlarged in Fig. 2(c) and (d). The $T_c^{onset}$ is regarded as the crossing point of the fitting lines for resistivity in the normal state near the transition and the drop area during the transition. The $T_c^{zero}$ is estimated as the crossing point of the line for zero resistivity and the $\rho$-$T$ curve. Superconductivity does not appear between $x$ = 0.0 and 0.4. Above $x$ = 0.45, the onset of the superconducting transition is observed for all the samples. Zero resistivity is observed between $x$ = 0.45 and 0.7. The temperature dependence of resistivity for $CeO_{1-x}F_xBiS_2$ ($x$ = 0 - 1.0) after high-pressure annealing is shown in Fig. 3(a).

All the samples show a semiconducting behavior in the normal state. The samples above $x$ = 0.3 exhibits the onset of the superconducting transition, as shown in Fig. 3(b). A $T_c^{onset}$ is observed around 3 K for samples where $x$ = 0.3, 0.4, 0.45, and 1.0. The sample with $x$ = 0.6 shows the onset of superconducting transition around 6 K. For $x$ = 0.7 and 0.9, $T_c^{onset}$ is observed around 8 K. $T_c^{zero}$ is not observed between $x$ < 0.6 in the temperature range down to 2 K. For 0.6 ≤ $x$ ≤ 0.9, bulk superconductivity is observed. Among them, $CeO_{0.3}F_{0.7}BiS_2$ shows the maximum $T_c^{onset}$ of 7.8 K. The fact that bulk superconductivity with higher $T_c$ appears for $x$ ≥ 0.6 seems to be consistent with the observation of structural change revealed in Fig. 1(c).

The temperature dependence of magnetic susceptibility for the AG samples of the $CeO_{1-x}F_xBiS_2$ is shown in Fig. 4(a) and (b). Paramagnetic behavior is observed for samples where $x$ = 0.0 - 0.4. A superconducting transition appears around 2.5 K for samples where $x$ = 0.45, 0.5 and o.6. In these samples, the magnetic susceptibility increases around 4.5 K with decreasing temperature. Furthermore, the $x$ = 0.6 sample exhibits two anomalies at 4.5 K and 7.5 K. Above $x$ = 0.7, the magnetic susceptibility largely increases at 7.5K. The increase of magnetic susceptibility suggests that two ferromagnetic phases with respective magnetic transition temperatures of 4.5 K and 7.5 K exist. Magnetic susceptibility measurements for HP samples are presented in Fig. 5(a) and (b). After high pressure annealing, superconductivity around 6 K appears when $x$ = 0.7 and 0.9. In this region, ferromagnetism at 7.5 K is dominant. $CeO_{0.3}F_{0.7}BiS_2$ shows the strongest diamagnetic signal among all the samples, while exhibiting ferromagnetic ordering below 7.5 K.

We summarize the F concentration dependence of ferromagnetism and the superconducting transition temperature in the phase diagram of Fig. 6(a) and (b) to understand the physical properties of the $CeO_{1-x}F_xBiS_2$ system. From Fig. 6(a), the ferromagnetic phases observed at 4.5 K and 7.5 K are given the terms lower temperature phase and higher temperature phase respectively. The lower temperature phase (4.5 K) exists between $x$ = 0.45 and 0.6, and coexists with the higher temperature phase (7.5 K) at $x$ = 0.6. Above $x$ = 0.6, the higher temperature phase (7.5 K) can be observed. In Fig. 6(b), the AG samples show superconductivity from $x$ = 0.45 to 1.0. After high-pressure annealing, superconductivity appears in a large region between $x$ = 0.3 and 1.0. The $T_c$ of the HP samples is obviously higher than that of the AG samples.

## Discussion

In this section, we will discuss the interesting points of $CeO_{1-x}F_xBiS_2$. We found that ferromagnetism and superconductivity simultaneously develop with systematic F doping. In particular, two ferromagnetic phases with respective transition temperatures of 4.5 K and 7.5 K were induced with increasing F concentration. This result indicates that carriers generated by F doping are provided to, not only the superconducting $BiS_2$ layers, but also the blocking

CeO layers. Thus, we suggest that the observed two ferromagnetic phases are a result of the change in the valence of the Ce ion with changing the F concentration. We found that bulk superconductivity of $CeO_{1-x}F_xBiS_2$ could be induced with $x \geq 0.6$ in the HP samples. Interestingly, the F concentration where bulk superconductivity appears corresponds to the F concentration where ferromagnetism with a transition temperature of 7.5 K. In this regard, we assume that superconductivity and ferromagnetism are linked to each other in the $CeO_{1-x}F_xBiS_2$. In fact, the region with F concentration level of $x \geq 0.7$ corresponds with the F concentration where the XRD peaks show the obvious change from the tetragonal phase with lower F concentration. Above $x = 0.7$, the crystal structure is possibly optimized for the appearance of both ferromagnetism and bulk superconductivity due to high F concentration. In order to reveal this simultaneous development of ferromagnetism and bulk superconductivity, it is quite important to carry out crystal structure analysis and measurements sensitive to the magnetic ordering and the valence state of the Ce ion using single crystals of $CeO_{1-x}F_xBiS_2$.

**Conclusion**

We found that the coexistence of bulk superconductivity and ferromagnetism can be realized in the $BiS_2$-based superconductor $CeO_{1-x}F_xBiS_2$ prepared by high-pressure annealing. In the as-grown samples, weak superconductivity phase and two kinds of ferromagnetic orderings with transition temperatures of 4.5 K and 7.5 K are induced through the continuous substitution of F. This result indicates that carriers generated by the F substitution contribute to both the superconducting and blocking layers. Furthermore, we found that $T_c$ can be increased from 3 K to 8 K by high-pressure annealing at $x$ = 0.7 and 0.9. These samples show strong diamagnetism while keeping the ferromagnetic ordering. Therefore, bulk superconductivity and ferromagnetism intrinsically coexist in $CeO_{1-x}F_xBiS_2$ prepared with high-pressure annealing. The enhancement of $T_c$ is achieved around $x$ = 0.7, where the higher temperature ferromagnetic phase (7.5 K) becomes dominant. This fact implies that the appearance of the higher $T_c$ is linked to the evolution of ferromagnetism with the higher magnetic transition temperature of 7.5 K. The $CeO_{1-x}F_xBiS_2$ system is a potential system for discussing the correlation between superconductivity and ferromagnetism, and for further applications in superconductivity.

**Acknowledgements**

This work was partly supported by a Grant-in-Aid for Scientific Research from the Ministry of Education, Culture, Sports, Science and Technology (KAKENHI), Strategic International Collaborative Research Program (SICORP-EU-Japan), and Japan Science and Technology Agency and “Funding program for World-Leading Innovative R&D on Science Technology (FIRST) Program”.

## References

[1] J. G. Bednorz, and K. Müller: Z. Physik B Condensed Matter 64 (1986) 189.

[2] M. K. Wu, J. R. Ashburn, C. J. Torng, P. H. Hor, R. L. Meng, L. Gao, Z. J. Huang, Y. Q. Wang, and C. W. Chu: Phys. Rev. Lett. **58** (1987) 908.

[3] H. Maeda, Y. TAnaka, and T. Asano: Jpn. J. Appl. Phys. **27** (1988) L209.

[4] A.Schilling M. Cantoni, J. D. Guo, and H. R. Ott: Nature **363** (1993) 56.

[5] Y. Kamihara, T. Watanabe, M. Hirano, and H. Hosono: J. Am. Chem. Soc. **130** (2008) 3296.

[6] X. H.Chen T. Wu, G. Wu, R. H. Liu, H. Chen, and D. F. Fang: Nature **453** (2008) 761.

[7] Z. A. Ren, W. Lu, J. Yang, W. Yi, X. L. Shen, C. Zheng, G. C. Che, X. L. Dong, L. L. Sun, F. Zhou, and Z. X. Zhao: Chinese Phys. Lett. **25** (2008) 2215.

[8] M. Rotter, M. Tegel, and D. Johrendt: Phys. Rev. Lett. **101** (2008) 107006.

[9] X. C. Wang, Q. Q. Liu, Y. X. Lv, W. B. Gao, L. X. Yang, R. C. Yu, F. Y. Li, and C. Q. Jin, Solid State Commun. **148** (2008) 538.

[10] F. C. Hsu, J. Y. Luo, K. W. The, T. K. Chen, T. W. Huang, P. M. Wu, Y. C. Lee, Y. L. Huang, Y. Y. Chu, D. C. Yan, and M. K. Wu: Proc. Natl. Acad. Sci. U. S. A. **105** (2008) 14262.

[11] K. W. Yeh T. W. Huang, Y. L. Huang, T. K. Chen, F. C. Hsu, P. M. Wu, Y. C. Lee, Y. Y. Chu, C. L. Chen, J. Y. Luo, D. C. Yan, and M. K. Wu: EPL **84** (2008) 37002.

[12] Y. Mizuguch, F. Tomioka, T. Yamaguchi, and Y. Takano: Appl. Phys. Lett. **94** (2009) 012503.

[13] J. Guo, S. Jin, G. Wang, K. Zhu, T. Zhou, M. He, and X. Chen: Phys. Rev. B **82** (2010) 180520.

[14] S. Jiang, H. Xing, G. Xuan, Z. Ren, C. Wang, Z. A. Xu, and G. Cao, Phys. Rev. B 80, (2009) 184514.

[15] H. S. Jeevan, Z. Hossain, D. Kasinathan, H. Rosner, C. Geibel, and P. Gegenwart, Phys. Rev. B 78, (2008) 092406.

[16] H. S. Jeevan, D. Kasinathan, H. Rosner, amd P. Gegenwart: Phys. Rev. B **83** (2011) 054511.

[17] G. Cao, S. Xu, Z. Ren, S. Jiang, C. Feng, and Z. A. Xu, J. Phys. Condens. Matter **23**, (2011) 464204.

[18] Z. Ren, Q. Tao, S. Jiang, C. Feng, C. Wang, J. Dai, G. Cao, and Z. Xu, Phys. Rev. Lett. **102**, (2009) 137002.

[19] W. H. Jiao, Q. Tao, J. K. Bao, Y. L. Sun, C. M. Feng, Z. A. Xu, I. Nowik, I. Felner, and G. H. Cao, Europhys. Lett. **95**, (2011) 67007.

[20] T. Terashima, M. Kimata, H. Satsukawa, A. Harada, K. Hazama, S. Uji, H. S. Suzuki, T. Matsumoto, and K. Murata, J. Phys. Soc. Jpn. **78**, (2009) 083701.

[21] K. Matsubayashi, K. Munakata, M. Isobe, N. Katayama, K. Ohgushi, Y. Ueda, Y.

Uwatoko, N. Kawamura, M. Mizumaki, N. Ishimatsu, M. Hedo, and I. Umehara, Phys. Rev. B **84**, (2011) 024502.

[22] G. F. Chen, Z. Li, D. Wu, G. Li, W. Z. Hu, J. Dong, P. Zheng, J. L. Luo, and N. L. Wang, Phys. Rev. Lett. **100**, (2008) 247002.

[23] Y. Luo, Y. Li, S. Jiang, J. Dai, G. Cao, and Z. Xu, Phys. Rev. B **81**, (2010) 134422.

[24] H. Kotegawa, Y. Tokunaga, Y. Araki1, G.-q. Zheng, Y. Kitaoka, K. Tokiwa, K. Ito, T. Watanabe, A. Iyo, Y. Tanaka, and H. Ihara, Phys. Rev. B **69**, (2004) 014501.

[25] H. Mukuda, M. Abe, S. Shimizu, Y. Kitaoka, A. Iyo, Y. Kodama1, H. Kito, Y. Tanaka, K. Tokiwa and T. Watanabe, J. Phys. Soc. Jpn. **75** (2006) 123702.

[26] H. Mukuda, M. Abe, Y. Araki, Y. Kitaoka, K. Tokiwa, T. Watanabe, A. Iyo, H. Kito, and Y. Tanaka, Phys. Rev. Lett. 96, 087001(2006).

[27] S. Shimizu, H. Mukuda, Y. Kitaoka, H. Kito, Y. Kodama, P. M. Shirage, and A. Iyo, J. Phys. Soc. Jpn. **78** (2009) 064705.

[28] S. Shimizu, S. Tabata, H. Mukuda, Y. Kitaoka, P. M. Shirage, H. Kito, and A. Iyo, J. Phys. Soc. Jpn. **80** (2011) 043706.

[29] Y. Mizuguchi, H. Fujihisa, Y. Gotoh, K. Suzuki, H. Usui, K. Kuroki, S. Demura, Y. Takano, H. Izawa, and O. Miura: Phys. Rev. B **86** (2012) 220510(R).

[30] Y. Mizuguchi, S. Demura, K. Deguchi, Y. Takano, H. Fujihisa, Y. Gotoh, H. Izawa, and O. Miura: J. Phys. Soc. Jpn. **81** (2012) 114725.

[31] S. Demura, Y. Mizuguchi, K. Deguchi, H. Okazaki, H. Hara, T. Watanabe, S. J. Denholme, M. Fujioka, T. Ozaki, H. Fujihisa, Y. Gotoh, O. Miura, T. Yamaguchi, H. Takeya, Y. Takano: J. Phys. Soc. Jpn. **82** (2013) 033708.

[32] H. Lei, K. Wang, M. Abeykoon, E. S. Bozin, C. Petrovic, arXiv:1208.3189

[33] J. Xing, S. Li, X. Ding, H. Yang, Hai-Hu Wen, Phys. Rev. B **86**, (2012) 214518

[34] R. Jha, A. Kumar, S. K. Singh, V. P. S. Awana, J. Sup. and Novel Mag. **26**, (2013) 499.

[35] D. Yazici, K. Huang, B. D. White, A. H. Chang, A. J. Friedman, M. B. Maple, Philosophical Magazine **93**, (2012) 673.

[36] D. Yazici, K. Huang, B. D. White, I. Jeon, V. W. Burnett, A. J. Friedman, I. K. Lum, M. Nallaiyan, S. Spagna, M. B. Maple, arXiv:1303.6216

[37] X. Lin, X. Ni, B. Chen, X. Xu, X. Yang, J. Dai, Y. Li, X. Yang, Y. Luo, Q. Tao, G. Cao, Zhuan Xu, Phys. Rev. B **87**, (2013) 020504(R).

[38] M. Nagao, S. Demura, K. Deguchi, A. Miura, S. Watauchi, T. Takei, Y. Takano, N. Kumada and I. Tanaka, J. Phys. Soc. Jpn. **81**. (2012) 103702.

[39] M. Nagao, A. Miura, S. Demura, K. Deguchi, S. Watauchi, T. Takei, Y. Takano, N. Kumada and I. Tanaka, Solid State. Commun. **178** (2014) 33.

[40] J. Liu. D. Fang, Z. Wang. J. Xing, Z. Du, X. Zhu, H. Yang, Hai-Hu Wen, arXiv:1310.0377.

[41] H. Kotegawa, Y. Tomita, H. Tou, H. Izawa, Y. Mizuguchi, O. Miura, S. Demura, K.

Deguchi, Y. Takano, J. Phys. Soc. Jpn. **81** (2012) 103702.

[42] G. K. Selvan, M. Kanagaraj, S. E. Muthu, R. Jha, V. P. S. Awana, S. Arumugam, Phys. Stat. Sol. Rapid Res. Lett. (2013).

[43] C. T. Wolowiec, D. Yazici, B. D. White, K. Huang and M. B. Maple, arXiv:1307.4157.

[44] G. Kalai Selvan, M. Kanagaraj, Rajveer Jha, V. P. S. Awana, S. Arumugam, arXiv:1307.4877.

[45] C. T. Wolowiec, B. D. White, I. Jeon, D. Yazici, K. Huang, M. B. Maple, arXiv:1308.1072.

[46] T. Tomita, M. Ebata, H. Soeda, H. Takahashi, H. Fujihisa, Y. Gotoh, Y. Mizuguchi, H. Izawa, O. Miura, S. Demura, K. Deguchi, Y. Takano, arxiv:1309. 4250.

[47] K. Deguchi, Y. Mizuguchi, S. Demura, H. Hara, T. Watanabe, S. J. Denholme, M. Fujioka, H. Okazaki, T. Ozaki, H. Takeya, T. Yamaguchi, O. Miura, and Y. Takano: Europhys. Lett. **101** (2013) 17004.

**Figure caption**

Fig. 1

(a, b) X-ray diffraction patterns for $CeO_{1-x}F_xBiS_2$ ($x$ = 0 - 1.0) of as-grown and high-pressure samples. Filled circles indicate the peaks of the impurity phases. (c) Enlarged figure of (110), (114) and (200) peaks for AS samples. Black arrows denote the new peaks emerging by high F substitution. (d, e) The nominal $x$ dependence of the lattice constants $a$ and $c$ for as-grown and high-pressure samples, respectively.

Fig. 2

(a, b) The temperature dependence of resistivity for $CeO_{1-x}F_xBiS_2$ ($x$ = 0 - 1.0) of as-grown samples between 300 and 2 K. (c, d) The enlarged picture of resistivity between 10 and 2 K.

Fig. 3

(a) The temperature dependence of resistivity for $CeO_{1-x}F_xBiS_2$ ($x$ = 0 - 1.0) after the high-pressure annealing between 300 and 2 K. (b) The enlarged picture of resistivity between 10 and 2 K.

Fig. 4

(a, b) The temperature dependence of the magnetic susceptibility for as-grown $CeO_{1-x}F_xBiS_2$ ($x$ = 0 - 1.0).

Fig. 5

(a, b) The temperature dependence of the magnetic susceptibility for $CeO_{1-x}F_xBiS_2$ ($x$ = 0 -

1.0) after the high-pressure annealing.

Fig. 6

(a, b) The nominal F concentration *x* - ferromagnetic and superconducting transition temperature phase diagram for $CeO_{1-x}F_xBiS_2$ ($x$ = 0 - 1.0), respectively.

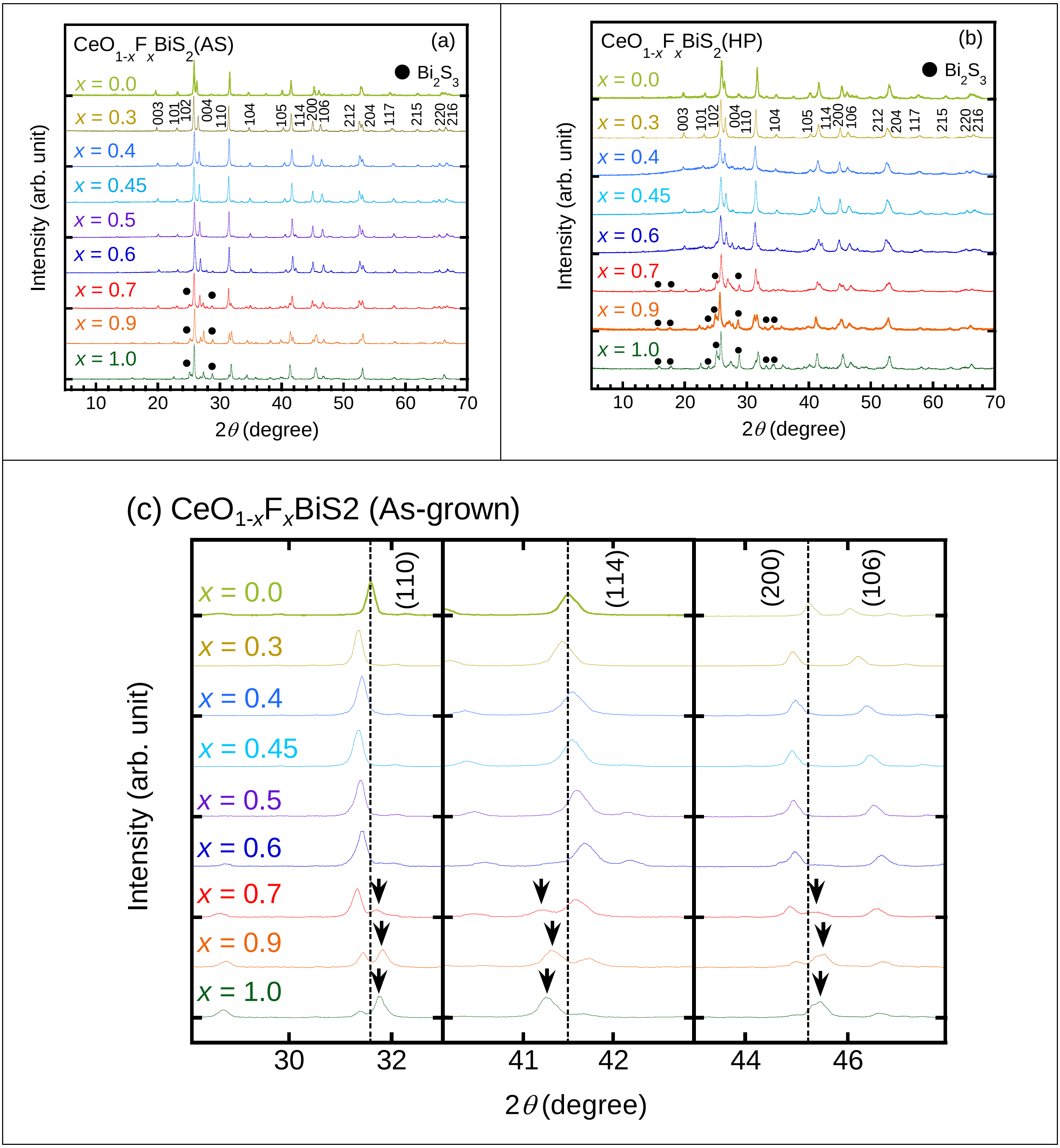

Fig. 1(a, b, c)

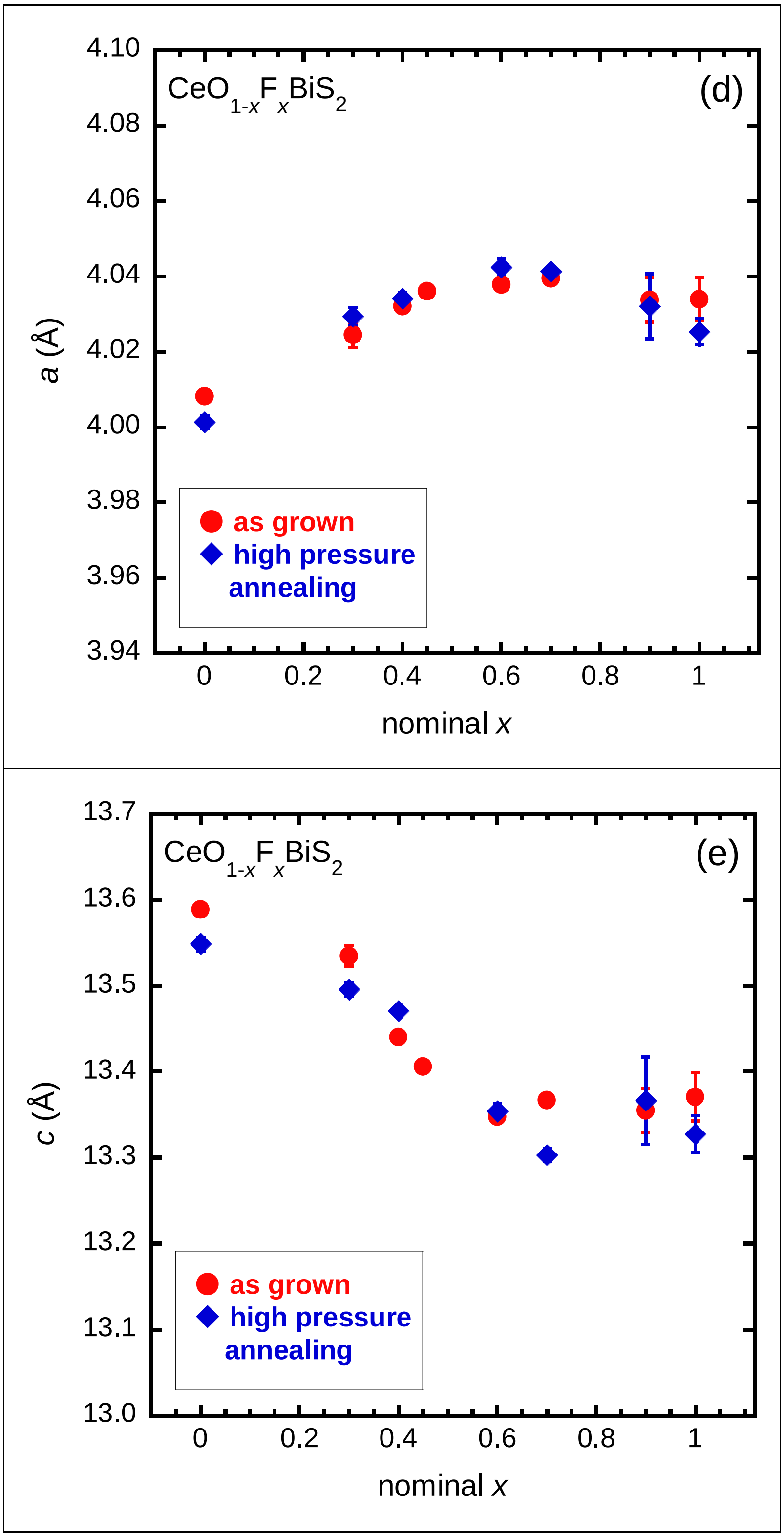

Fig. 1(d , e)

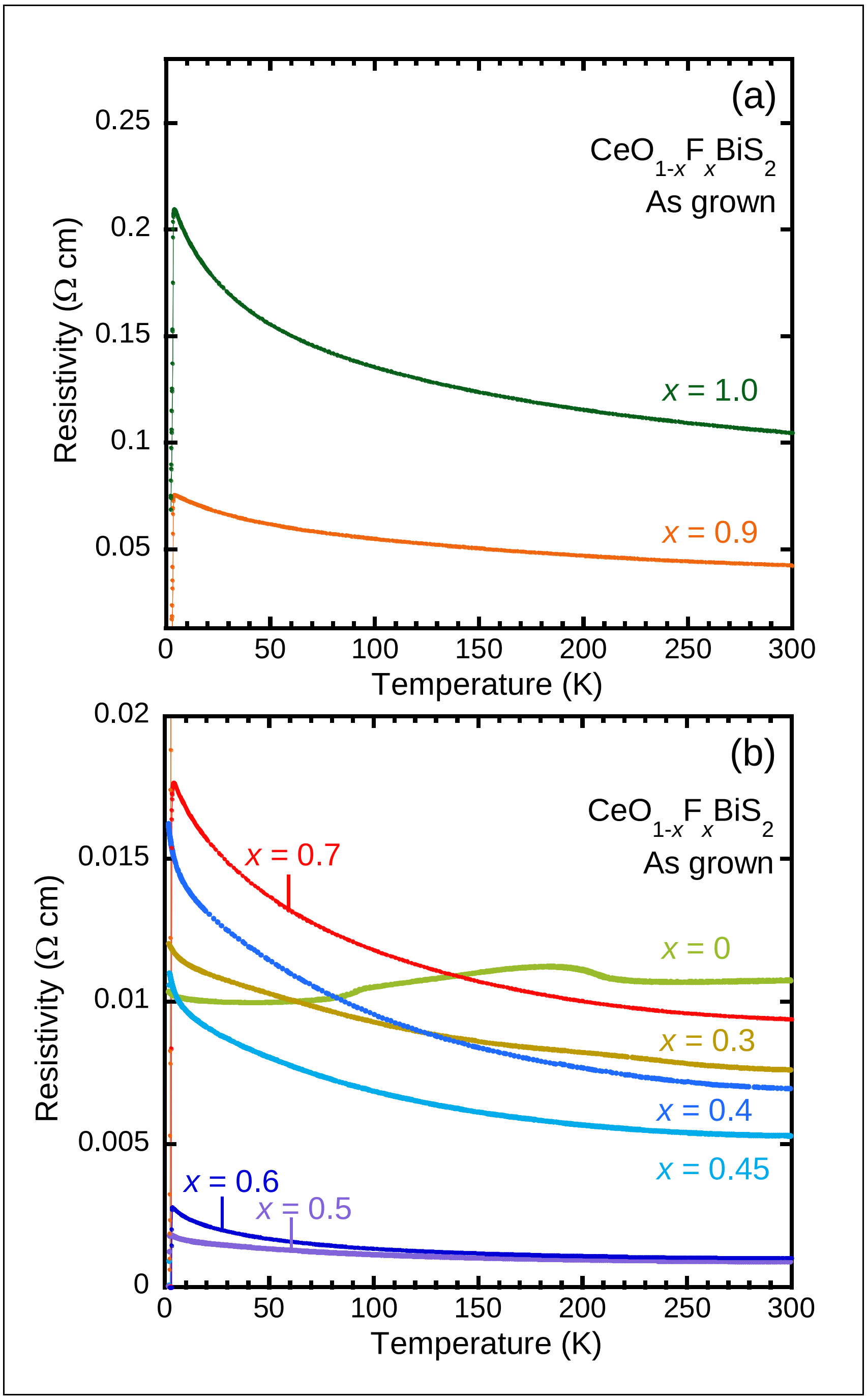

Fig. 2(a,b)

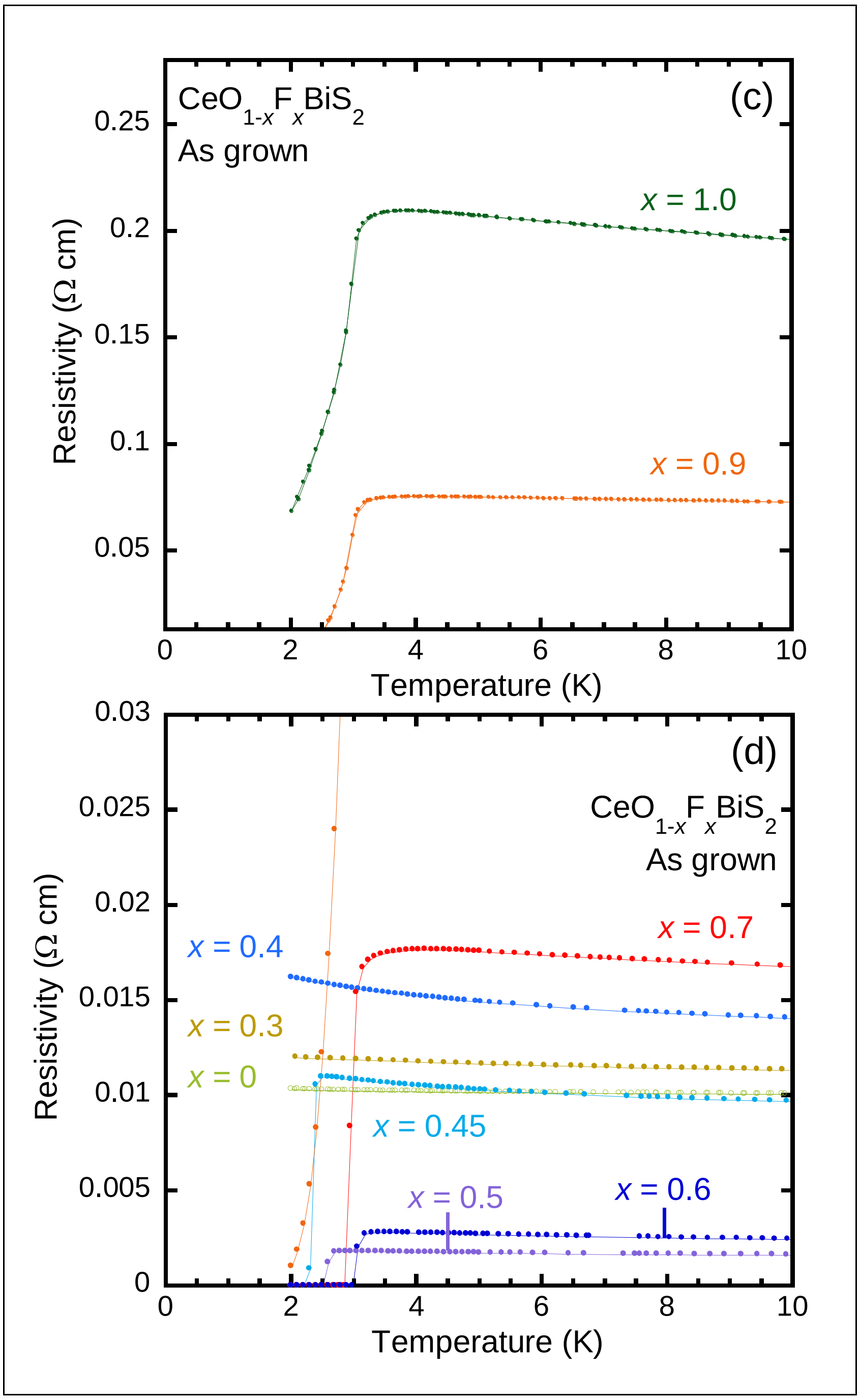

Fig. 2(c,d)

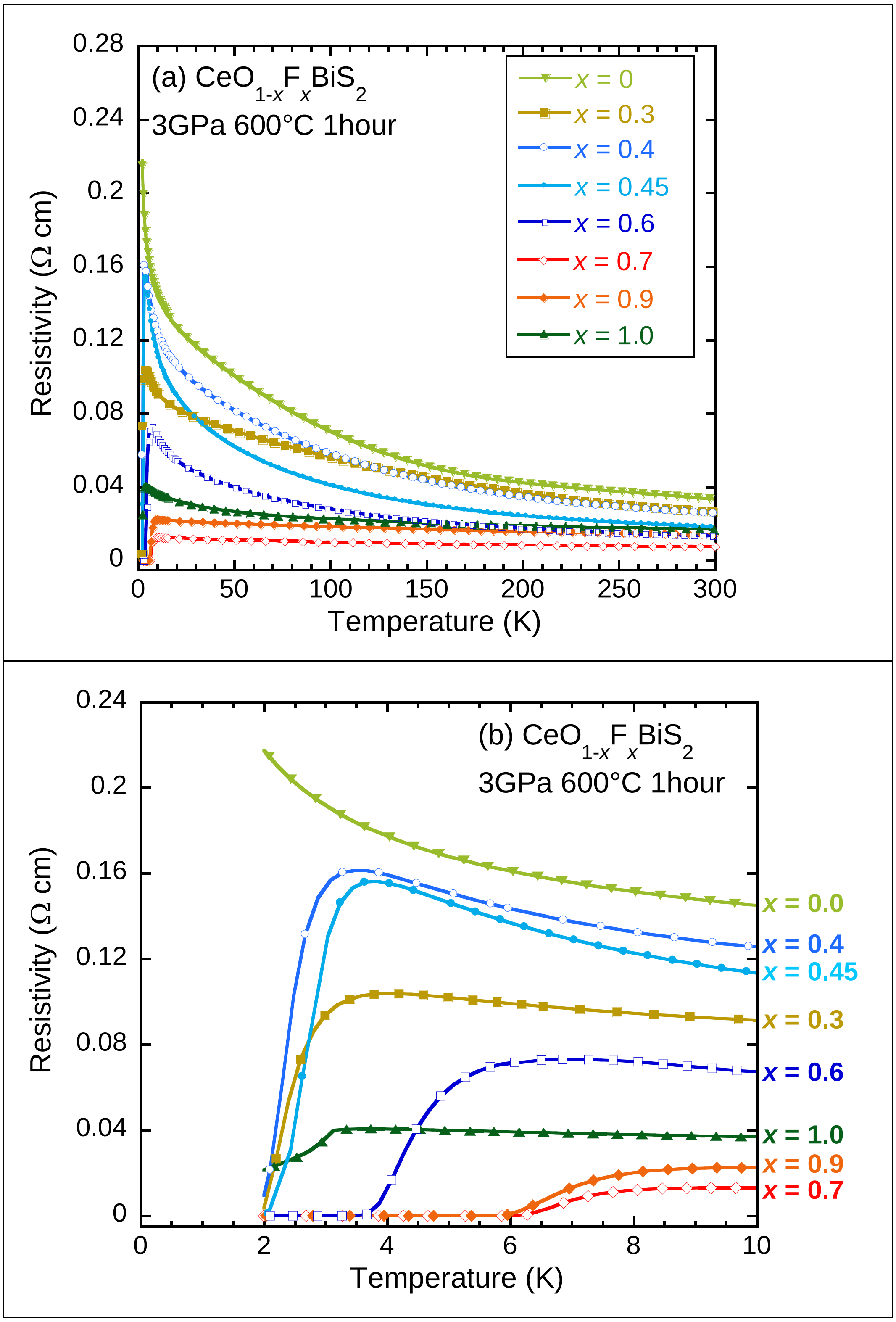

Fig. 3(a, b)

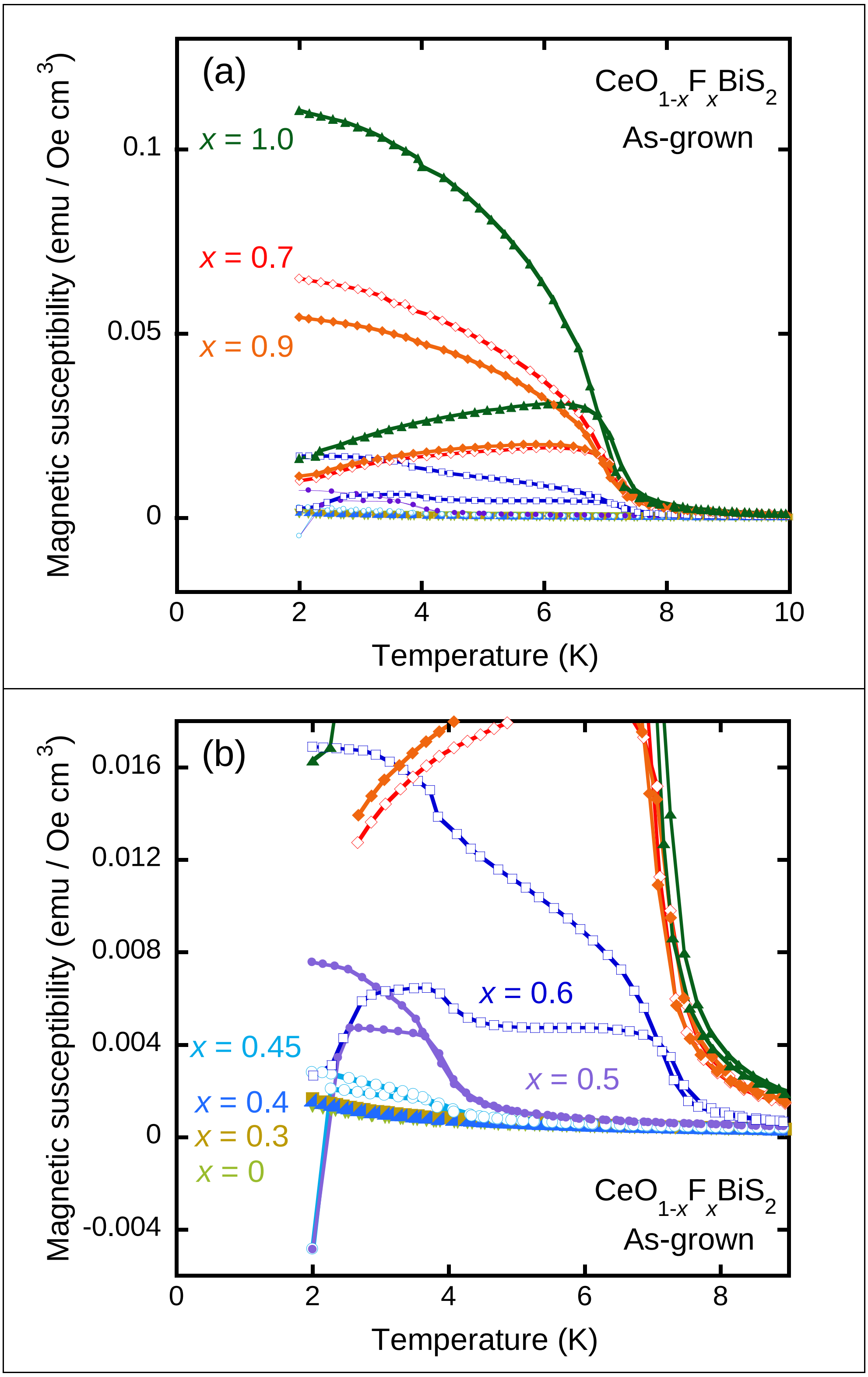

Fig. 4(a) , (b)

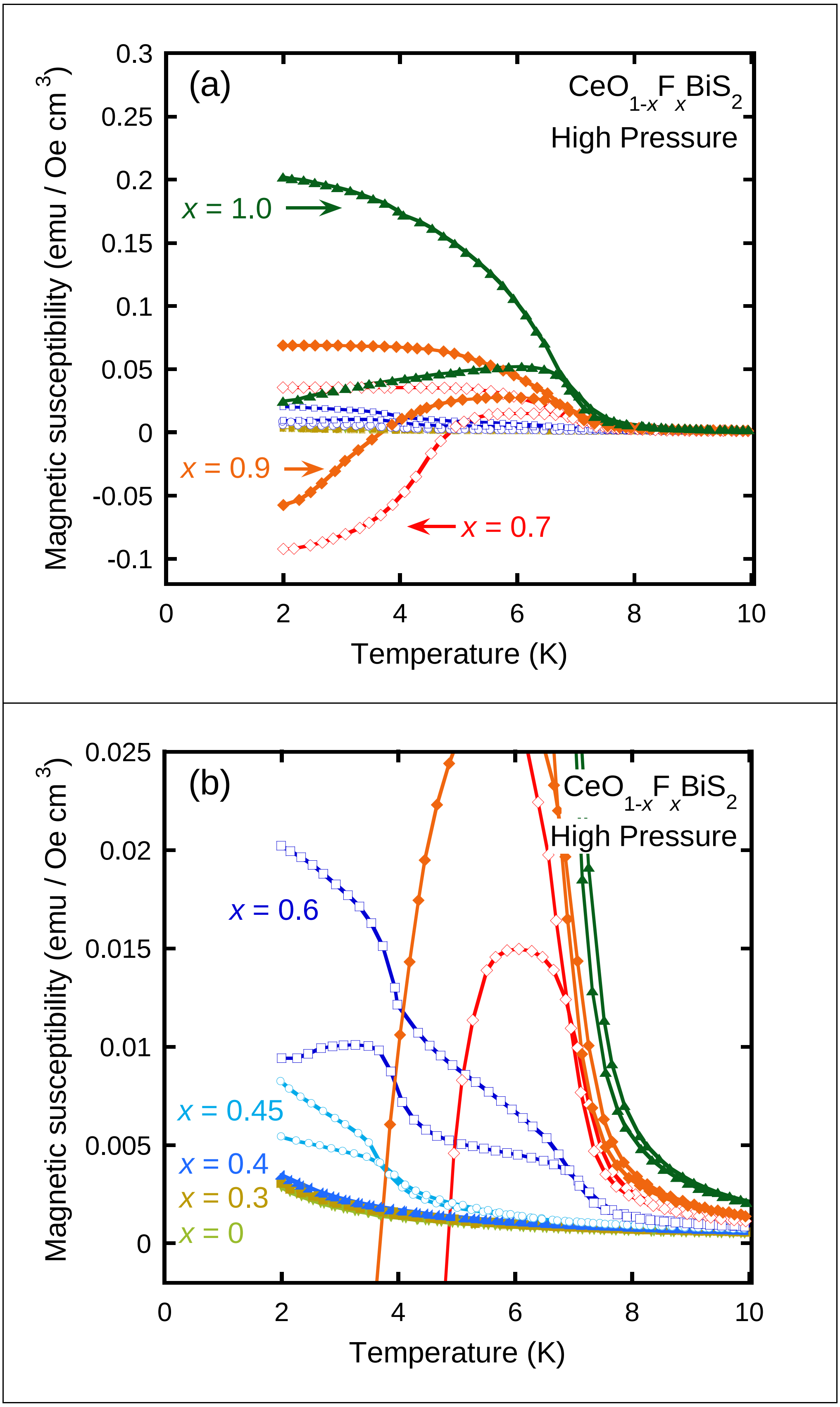

Fig. 5(a, b)

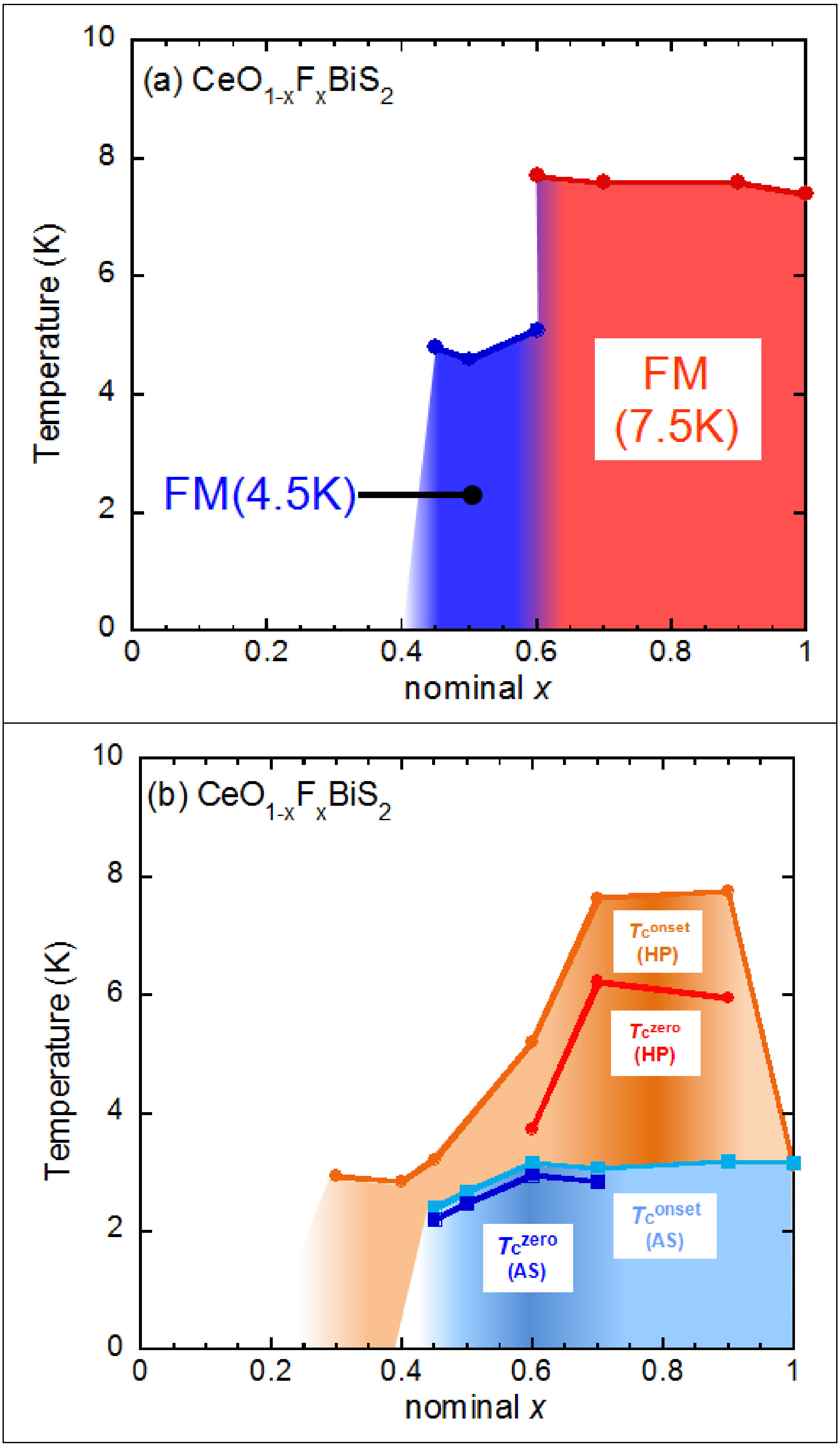

Fig. 6(a, b)